\documentclass[prx, twocolumn, superscriptaddress, longbibliography]{revtex4-1}
\usepackage{bm, amsmath, amsfonts, amssymb, ascmac, braket}
\usepackage[dvipdfmx]{graphicx}
\usepackage{float}
\usepackage{color}

\newcommand{\ii}{\text{i}}

\begin{document}

\title{Non-Bloch band theory of non-Hermitian Hamiltonians in the symplectic class}

\author{Kohei Kawabata}
	\email{kawabata@cat.phys.s.u-tokyo.ac.jp}
	\affiliation{Department of Physics, University of Tokyo, 7-3-1 Hongo, Bunkyo-ku, Tokyo 113-0033, Japan}
	
\author{Nobuyuki Okuma}
	\affiliation{Yukawa Institute for Theoretical Physics, Kyoto University, Kyoto 606-8502, Japan}
	
\author{Masatoshi Sato}
	\affiliation{Yukawa Institute for Theoretical Physics, Kyoto University, Kyoto 606-8502, Japan}

\date{\today}

\begin{abstract} 
Non-Hermitian Hamiltonians are generally sensitive to boundary conditions, and their spectra and wave functions under open boundary conditions are not necessarily predicted by the Bloch band theory for periodic boundary conditions. To elucidate such a non-Bloch feature, recent works have developed a non-Bloch band theory that works even under arbitrary boundary conditions. Here, it is demonstrated that the standard non-Bloch band theory breaks down in the symplectic class, in which non-Hermitian Hamiltonians exhibit Kramers degeneracy because of reciprocity. Instead, a modified non-Bloch band theory for the symplectic class is developed in a general manner, as well as illustrative examples. This nonstandard non-Bloch band theory underlies the $\mathbb{Z}_{2}$ non-Hermitian skin effect protected by reciprocity.
\end{abstract}

\maketitle

\section{Introduction}
	\label{sec: introduction}

Topology plays a central role in contemporary physics. In particular, it describes a variety of phases of matter that cannot be described by spontaneous symmetry breaking~\cite{Kane-review, Zhang-review}. Such topological phases are ubiquitous in insulators~\cite{SSH-79, Haldane-88, Kane-Mele-05-QSH, *Kane-Mele-05-Z2} and superconductors~\cite{Read-Green-00, Kitaev-01}, as well as semimetals~\cite{Wan-11}, all of which are classified according to symmetry~\cite{Schnyder-08, *Ryu-10, Kitaev-09, Schnyder-Ryu-review}. A signature of topology manifests itself as the bulk-boundary correspondence: nontrivial bulk topology of Bloch Hamiltonians results in the emergence of anomalous boundary states. For example, zero modes appear at the two ends of one-dimensional systems~\cite{SSH-79, Kitaev-01}, and chiral~\cite{Haldane-88, Read-Green-00} or helical~\cite{Kane-Mele-05-QSH, *Kane-Mele-05-Z2} gapless modes appear at the edges of two-dimensional systems. Remarkably, certain topological phases and their anomalous boundary modes are protected by symmetry. As a prime example, topological phases in quantum spin Hall insulators~\cite{Kane-Mele-05-QSH, *Kane-Mele-05-Z2} are protected by time-reversal symmetry (reciprocity). Time-reversal-invariant (reciprocal) Hamiltonians $H$ respect
\begin{equation}
\mathcal{T} H^{T} \mathcal{T}^{-1} = H,\quad
\mathcal{T}\mathcal{T}^{*} = -1
	\label{eq: symplectic symmetry}
\end{equation}
with a unitary matrix $\mathcal{T}$ (i.e., $\mathcal{T}\mathcal{T}^{\dag} = \mathcal{T}^{\dag}\mathcal{T} = 1$), and are defined to belong to the symplectic class (class AII). An important consequence of this symmetry is Kramers degeneracy, which ensures the $\mathbb{Z}_{2}$ topological phase and the helical edge states~\cite{Kane-Mele-05-QSH, *Kane-Mele-05-Z2}.

Despite the enormous success, the existing framework of topological phases was confined to Hermitian systems. Nevertheless, non-Hermiticity appears, for example, in a variety of nonequilibrium open systems as a consequence of nonconservation of energy or particles~\cite{Konotop-review, Christodoulides-review}. To understand the role of topology in non-Hermitian systems, topological characterization of non-Hermitian systems~\cite{Ota-review, Bergholtz-review} has recently been developed both in theory~\cite{Rudner-09, Sato-11, *Esaki-11, Hu-11, Schomerus-13, Lee-16, Leykam-17, Xu-17, Shen-18, *Kozii-17, Gong-18, YW-18-SSH, *YSW-18-Chern, Kawabata-19, Kunst-18, KSU-18, McDonald-18, Lee-19, Jin-19, Liu-19, Lee-Li-Gong-19, Kunst-19, KSUS-19, ZL-19, Herviou-19, Zirnstein-19, Borgnia-19, KBS-19, YM-19, McClarty-19, Okuma-19, Song-19-Lindblad, Song-19-real, Bergholtz-19, Lee-Vishwanath-19, Rui-19, Schomerus-20, Imura-19, Herviou-19-ES, Chang-19, Zhang-20, Zhang-19, *Yang-19, OKSS-20, Longhi-20, Wang-20, Koch-20, Yoshida-19-mirror, Yokomizo-20, Li-20} and experiments~\cite{Poli-15, Zeuner-15, Zhen-15, *Zhou-18, Weimann-17, Xiao-17, St-Jean-17, Bahari-17, Harari-18, *Bandres-18, Zhao-19, Brandenbourger-19-skin-exp, *Ghatak-19-skin-exp, Helbig-19-skin-exp, *Hofmann-19-skin-exp, Xiao-19-skin-exp, Weidemann-20-skin-exp}. On the basis of the 38-fold internal symmetry in non-Hermitian physics~\cite{Bernard-LeClair-02, KSUS-19}, topological classification of non-Hermitian systems was established~\cite{Shen-18, Gong-18, KSUS-19, ZL-19, KBS-19}, which predicts a number of non-Hermitian topological phases that have no analog in Hermitian systems. 

Furthermore, non-Hermiticity is found to alter the nature of the bulk-boundary correspondence~\cite{Lee-16}. This breakdown arises from the extreme sensitivity of non-Hermitian systems to boundary conditions, which is called the non-Hermitian skin effect~\cite{YW-18-SSH, *YSW-18-Chern, Kunst-18}. In fact, spectra and wave functions of non-Hermitian systems under open boundary conditions can be strikingly different from those under periodic boundary conditions, only the latter of which are predicted by the Bloch band theory. To elucidate such a non-Bloch feature of non-Hermitian systems, recent works have developed a non-Bloch band theory that works even under arbitrary boundary conditions~\cite{YW-18-SSH, *YSW-18-Chern, Kunst-18, Lee-19, YM-19, Kunst-19, Zhang-19, OKSS-20}. Reference~\cite{Lee-16} numerically investigated a non-Hermitian extension of the Su-Schrieffer-Heeger model~\cite{SSH-79} with asymmetric hopping, which is a prototypical example that exhibits the non-Hermitian skin effect. Providing the exact solution to this model, Ref.~\cite{YW-18-SSH} showed a clear understanding about the non-Hermitian skin effect and the non-Bloch bulk-boundary correspondence. Reference~\cite{YM-19} further generalized this result and gave a non-Bloch band theory in a general manner, which is summarized as follows:

\medskip
\begin{itembox}[1]{Non-Bloch band theory~\cite{YW-18-SSH, YM-19}}
Suppose $H \left( \beta \right)$ denotes a bulk Hamiltonian in one dimension with $\beta := e^{\ii k}$ and complex-valued wavenumbers $k \in \mathbb{C}$. Moreover, $\beta_{i}$'s ($i = 1, 2, \cdots, 2M$; $\left| \beta_{1} \right| \leq \left| \beta_{2} \right| \leq \cdots \leq \left| \beta_{2M} \right|$) denote the solutions to the characteristic equation $\det \left[ H \left( \beta \right) - E \right] = 0$ in terms of $\beta$ for the given eigenenergy $E \in \mathbb{C}$. Then, continuum bands are formed by $H \left( \beta \right)$ with the trajectory of $\beta_{M}$ and $\beta_{M+1}$ satisfying
\begin{equation}
\left| \beta_{M} \right| = \left| \beta_{M+1} \right|.
	\label{eq: Yao-Wang-Yokomizo-Murakami}
\end{equation}
\end{itembox}

\bigskip 
In Hermitian systems, we always have $\left| \beta \right| = 1$ and all the eigenstates for continuum bands are delocalized through the bulk. In the non-Hermitian case, by contrast, this is not always the case, and an enormous number of localized states can appear, which is a signature of the skin effect. Correspondingly, wavenumbers are complex valued and form a generalized Brillouin zone. The non-Bloch band theory correctly describes a number of non-Hermitian systems and their topology. Recent experimental observations confirmed it in mechanical metamaterials~\cite{Brandenbourger-19-skin-exp, *Ghatak-19-skin-exp}, electrical circuits~\cite{Helbig-19-skin-exp, *Hofmann-19-skin-exp}, quantum walk~\cite{Xiao-19-skin-exp}, and photonic lattices~\cite{Weidemann-20-skin-exp}. It may bring about phenomena and functionalities unique to non-Hermitian systems, some of which were recently explored~\cite{Gong-18, McDonald-18, Song-19-Lindblad, Schomerus-20, Longhi-20}. However, the validity of the non-Bloch band theory has been unclear in the presence of symmetry.

In this work, although the standard non-Bloch band theory~\cite{YW-18-SSH, YM-19} is applicable to generic non-Hermitian systems without symmetry, we demonstrate its breakdown in the symplectic class. For non-Hermitian Hamiltonians, the symplectic class (class $\text{AII}^{\dag}$ in Ref.~\cite{KSUS-19}) is defined by reciprocity given by Eq.~(\ref{eq: symplectic symmetry}). Because of this symmetry, Hamiltonians exhibit Kramers degeneracy even in non-Hermitian systems, leading to the breakdown of the standard non-Bloch band theory. Instead, we generally provide a modified condition for continuum bands in the symplectic class, summarized as follows:

\medskip
\begin{itembox}[1]{Non-Bloch band theory in the symplectic class}
When non-Hermitian Hamiltonians respect reciprocity in Eq.~(\ref{eq: symplectic symmetry}) and belong to the symplectic class, the solutions to the characteristic equation $\det \left[ H \left( \beta \right) - E \right]$ are generally denoted as
\begin{equation}
\beta_{1}, \beta_{2}, \cdots, \beta_{2M}; \beta_{2M}^{-1}, \beta_{2M-1}^{-1}, \cdots, \beta_{1}^{-1}
\end{equation}
with $\left| \beta_{1} \right| \leq \cdots \left| \beta_{2M} \right| \leq 1 \leq | \beta_{2M}^{-1} | \leq \cdots \leq | \beta_{1}^{-1} |$. Here, $\beta_{i}$ and $\beta_{i}^{-1}$ form a Kramers pair. Then, the condition for continuum bands is given as
\begin{equation}
\left| \beta_{2M-1} \right| = \left| \beta_{2M} \right|.
	\label{eq: Yao-Wang-Yokomizo-Murakami-symplectic}
\end{equation}
\end{itembox}

\bigskip 
Remarkably, the standard non-Bloch band theory~\cite{YW-18-SSH, YM-19} predicts Eq.~(\ref{eq: Yao-Wang-Yokomizo-Murakami}), i.e., $\left| \beta_{2M} \right| = | \beta_{2M}^{-1} |$, for continuum bands, but this is not the case in the symplectic class. The condition~(\ref{eq: Yao-Wang-Yokomizo-Murakami}) intuitively implies the interference between the non-Bloch waves with $\beta_{M}$ and $\beta_{M+1}$. In the symplectic class, however, the non-Bloch waves with $\beta_{2M}$ and $\beta_{2M}^{-1}$ cannot interfere with each other since they form a Kramers pair; instead, the non-Bloch waves with $\beta_{2M-1}$ and $\beta_{2M}$ interfere, replacing the condition~(\ref{eq: Yao-Wang-Yokomizo-Murakami}) with the condition~(\ref{eq: Yao-Wang-Yokomizo-Murakami-symplectic}). This nonstandard non-Bloch band theory underlies a new type of non-Hermitian skin effects protected by reciprocity~\cite{OKSS-20}.

More precisely, the conditions~(\ref{eq: Yao-Wang-Yokomizo-Murakami}) and (\ref{eq: Yao-Wang-Yokomizo-Murakami-symplectic}) are derived from boundary conditions. In the standard (symplectic) case, boundary conditions impose a constraint on $\beta_{i}$'s, which forms an $M$-th-order (a $2M$-th-order) algebraic equation in terms of $\beta_{1}^{L}, \beta_{2}^{L}, \cdots, \beta_{M}^{L}$ ($\beta_{1}^{L}, \cdots, \beta_{2M}^{L}, \beta_{2M}^{-L}, \cdots, \beta_{1}^{-L}$) with the system size $L$ [see Eq.~(\ref{eq: det-symplectic}) for the symplectic case]. In the standard case, because of the assumption $\left| \beta_{1} \right| \leq \left| \beta_{2} \right| \leq \cdots \leq \left| \beta_{2M} \right|$, the leading-order term includes $( \beta_{M+1} \beta_{M+2} \cdots \beta_{2M} )^{L}$ and the next-to-leading-order term includes $( \beta_{M} \beta_{M+2} \cdots \beta_{2M} )^{L}$, in general. To respect the constraint, these two terms should be comparable to each other for $L \to \infty$, which leads to Eq.~(\ref{eq: Yao-Wang-Yokomizo-Murakami}). In the symplectic case, by contrast, reciprocity forbids the appearance of the term proportional to $( \beta_{2M}^{-1} \beta_{2M-1}^{-1} \cdots \beta_{1}^{-1} )^{L}$, which should be dominant in the absence of symmetry (see Sec.~\ref{sec: non-Bloch-symplectic; general} for details). Consequently, the leading-order-term including $( \beta_{2M} \beta_{2M-1}^{-1} \beta_{2M-2}^{-1} \cdots \beta_{1}^{-1} )^{L}$ and the next-to-leading-order term including $( \beta_{2M-2}^{-1} \beta_{2M-3}^{-1} \cdots \beta_{1}^{-1} )^{L}$ should be comparable, which yields Eq.~(\ref{eq: Yao-Wang-Yokomizo-Murakami-symplectic}).

This work is organized as follows. In Sec.~\ref{sec: reciprocity}, we summarize the non-Bloch band theory and basic properties of reciprocity. In Sec.~\ref{sec: non-Bloch-symplectic}, we begin with a symplectic extension of the Hatano-Nelson model~\cite{Hatano-Nelson-96, *Hatano-Nelson-97, OKSS-20}, which is a prototypical non-Hermitian system in the symplectic class. Then, we generally demonstrate the non-Bloch band theory in the symplectic class. In Sec.~\ref{sec: other symmetry}, we also discuss similar modification of the non-Bloch band theory in other symmetry classes. We conclude this work in Sec.~\ref{sec: conclusion}.

\section{Reciprocity}
	\label{sec: reciprocity}
	
Reciprocity is one of the fundamental internal symmetry~\cite{Bernard-LeClair-02, KSUS-19}. There are two types of reciprocity according to the sign of the unitary matrix $\mathcal{T}$ (i.e., $\mathcal{T}\mathcal{T}^{*} = +1$ or $\mathcal{T}\mathcal{T}^{*} = -1$), one of which is defined by
\begin{equation}
\mathcal{T} H^{T} \mathcal{T}^{-1} = H,\quad
\mathcal{T}\mathcal{T}^{*} = +1,
	\label{eq: orthogonal symmetry}
\end{equation}
and the other of which is defined by Eq.~(\ref{eq: symplectic symmetry}). In Ref.~\cite{KSUS-19}, this symmetry is called $\text{TRS}^{\dag}$ since it is a Hermitian-conjugate counterpart of time-reversal symmetry (TRS), and non-Hermitian Hamiltonians with Eqs.~(\ref{eq: orthogonal symmetry}) and (\ref{eq: symplectic symmetry}) are defined to belong to classes $\text{AI}^{\dag}$ and $\text{AII}^{\dag}$, respectively. In this work, classes $\text{AI}^{\dag}$ and $\text{AII}^{\dag}$ are also called the orthogonal and symplectic classes, respectively, in a similar manner to the Hermitian case. Reciprocity appears in a variety of non-Hermitian systems. For example, time-reversal-invariant Hermitian Hamiltonians with gain or loss (i.e., complex onsite potential) respect it and belong to the orthogonal (symplectic) class in the absence (presence) of the spin degrees of freedom. In addition, it is relevant to open quantum systems described by the Lindblad master equation~\cite{Hamazaki-19, Lieu-19, Sa-19}. 

In this section, we describe basic properties of reciprocity relevant to the non-Bloch band theory. We begin with reviewing the non-Bloch band theory in Sec.~\ref{sec: standard non-Bloch}. In Sec.~\ref{sec: orthogonal reciprocity}, we investigate the orthogonal class and show the absence of skin effects due to the symmetry. We next investigate the symplectic class in Sec.~\ref{sec: symplectic reciprocity}. In this case, the symmetry gives rise to Kramers degeneracy and does not necessarily lead to the absence of skin effects.

\subsection{Non-Bloch band theory}
	\label{sec: standard non-Bloch}
	
In the following, we consider a generic non-Hermitian Hamiltonian in one dimension described by
\begin{equation}
\hat{H} = \sum_{n} \sum_{j=-l}^{l} \sum_{\mu, \nu=1}^{q} H_{j, \mu\nu} \hat{c}_{n+j, \mu}^{\dag} \hat{c}_{n, \nu},
	\label{eq: Hamiltonian-orthogonal}
\end{equation}
where $\hat{c}_{n, \mu}$ ($\hat{c}_{n, \mu}^{\dag}$) is the annihilation (creation) operator on site $n$, and $H_{j, \mu\nu}$ is the single-particle Hamiltonian. Moreover, $n$ describes the spatial degrees of freedom, $l$ describes the hopping range, and $\mu, \nu$ describe the internal degrees of freedom per unit cell. We assume translation invariance under periodic boundary conditions. Thus, $H_{j, \mu\nu}$ is independent of sites $n$ away from the edges. Because of the noninteracting (quadratic) nature of the Hamiltonian, diagonalization of the many-body Hamiltonian $\hat{H}$ reduces to diagonalization of the single-particle Hamiltonian $H$, whose elements are given by $H_{j, \mu\nu}$. Let $E \in \mathbb{C}$ be a complex eigenenergy of $H$ and $\ket{\phi}$ ($\ket{\chi}$) be the corresponding right (left) eigenstate~\cite{Brody-14}:
\begin{equation}
H \ket{\phi} = E \ket{\phi},\quad
H^{\dag} \ket{\chi} = E^{*} \ket{\chi}.
	\label{eq: eigenequation}
\end{equation}

Because of translation invariance of $H$ away from the edges, the eigenstates are given by a linear combination of fundamental solutions
\begin{equation}
\sum_{n=1}^{L} \beta^{n}_{i} \ket{n} \ket{\phi_{i}} 
:= \sum_{n=1}^{L} \sum_{\mu=1}^{q} \beta^{n}_{i} \phi_{\mu}^{(i)} \ket{n} \ket{\mu},\quad
\beta_{i}, \phi_{\mu}^{(i)} \in \mathbb{C},
	\label{eq: fundamental solution}
\end{equation}
where $L$ is the number of unit cells, $\ket{n}$ is a state localized at site $n$, and $\ket{\mu}$ is a state with the internal degree $\mu$. This wave function is delocalized through the bulk for $\left| \beta \right| = 1$, while it is localized around the edge $n=1$ ($n=L$) for $\left| \beta \right| < 1$ ($\left| \beta \right| > 1$). The corresponding bulk Hamiltonian is described by
\begin{equation}
H \left( \beta \right) := \sum_{j=-l}^{l} H_{j} \beta^{j},
	\label{eq: def-bulk-Hamiltonian}
\end{equation}
where $H_{j}$ is a $q \times q$ matrix defined by $\left( H_{j} \right)_{\mu\nu} := H_{j, \mu\nu}$ and satisfies
\begin{equation}
H \left( \beta_{i} \right) \ket{\phi_{i}} = E \ket{\phi_{i}}.
	\label{eq: H-beta-right}
\end{equation} 
In the presence of Hermiticity, Eq.~(\ref{eq: fundamental solution}) is just a plane wave and $H \left( \beta \right)$ is a conventional Bloch Hamiltonian because of $\left| \beta \right| = 1$. The possible $\beta_{i}$'s for given $E$ are determined by the characteristic equation
\begin{equation}
\det \left[ H \left( \beta \right) - E \right] = 0,
	\label{eq: characteristic-standard}
\end{equation}
which is the $2lq$-th-order algebraic equation in terms of $\beta$. For these $\beta_{i}$'s ($i = 1, 2, \cdots, 2lq$), the right eigenstate $\ket{\phi}$ in real space can be represented as
\begin{equation}
\ket{\phi} = \sum_{i=1}^{2lq} \sum_{n=1}^{L} \beta^{n}_{i} \ket{n} \ket{\phi_{i}}.
\end{equation}

Remarkably, if the right eigenstate $\ket{\phi}$ is localized at one end, the corresponding left eigenstate $\ket{\chi}$ is localized at the other end. To see this property, we notice that $\ket{\chi}^{*}$ is, by definition, a right eigenstate of $H^{T}$ with the eigenenergy $E$. Then, let us consider transposition $H \rightarrow H^{T}$, which leads to the transformations
\begin{equation}
H_{j, \mu\nu} \rightarrow H_{-j, \nu\mu},
\end{equation}
and
\begin{eqnarray}
H \left( \beta \right) \rightarrow
\sum_{j=-l}^{l} H_{-j}^{T} \beta^{-j}
= \sum_{j=-l}^{l} H_{j}^{T} \beta^{-j}
= H^{T}\,( \beta^{-1} ).\qquad
	\label{eq: transposition-transformation}
\end{eqnarray}
This result implies that if $\beta$ satisfies Eq.~(\ref{eq: characteristic-standard}) for $H$, $\beta^{-1}$ satisfies Eq.~(\ref{eq: characteristic-standard}) for $H^{T}$, and vice versa. Recalling that $\ket{\chi}^{*}$ is a right eigenstate of $H^{T}$, we conclude that if $\ket{\phi}$ is localized at one end, $\ket{\chi}$ is localized at the other end, and vice versa. This also means that delocalization of $\ket{\phi}$ occurs simultaneously with delocalization of $\ket{\chi}$.

\subsection{Orthogonal class (class $\text{AI}^{\dag}$)}
	\label{sec: orthogonal reciprocity}

Reciprocity imposes some constraints on the eigenstates $\ket{\phi}$ and $\ket{\chi}$. In fact, in the orthogonal class, Eq.~(\ref{eq: orthogonal symmetry}) yields
\begin{eqnarray}
H \left( \mathcal{T} \ket{\chi}^{*} \right)
= \mathcal{T} H^{T} \ket{\chi}^{*}
= E \left( \mathcal{T} \ket{\chi}^{*} \right),
	\label{eq: reciprocity-eigenstates}
\end{eqnarray}
which means that $\mathcal{T} \ket{\chi}^{*}$ is also a right eigenstate with the eigenenergy $E$. The eigenenergies are, in general, not degenerate solely in the presence of Eq.~(\ref{eq: orthogonal symmetry}). Hence, the two right eigenstates are equivalent to each other, i.e.,
\begin{equation}
\ket{\phi} \propto \mathcal{T} \ket{\chi}^{*}.
	\label{eq: orthogonal-constraint}
\end{equation} 
Because of the relationship between $\ket{\phi}$ and $\ket{\chi}$ discussed in Sec.~\ref{sec: standard non-Bloch}, they are forbidden to be localized in the orthogonal class. In fact, if $\ket{\phi}$ were localized at one end, $\mathcal{T} \ket{\chi}^{*}$ would be localized at the other end, which contradicts Eq.~(\ref{eq: orthogonal-constraint}). Here, we use the fact that the internal-symmetry operation does not change the place at which eigenstates are localized. Thus, no skin effects appear in the orthogonal class, and this is why we call the symmetry in Eq.~(\ref{eq: orthogonal symmetry}) reciprocity.

The absence of skin effects in the orthogonal class can be derived also on the basis of the non-Bloch band theory~\cite{KSUS-19}. Since transposition transforms $H \left( \beta \right)$ to $H^{T}\,( \beta^{-1} )$ as shown in Eq.~(\ref{eq: transposition-transformation}), reciprocity for $H$ [i.e., Eq.~(\ref{eq: orthogonal symmetry})] imposes
\begin{equation}
\mathcal{T} H^{T} \left( \beta \right) \mathcal{T}^{-1} = H\,( \beta^{-1} ).
\end{equation} 
Then, when $\beta$ is a solution to the characteristic equation~(\ref{eq: characteristic-standard}), we have
\begin{eqnarray}
\det \left[ H\,( \beta^{-1} ) - E \right] 
&=& \det \left[ \mathcal{T} H^{T} \left( \beta \right) \mathcal{T}^{-1} - E \right] \nonumber \\
&=& \det \left[ H \left( \beta \right) - E \right] \nonumber \\
&=& 0,
	\label{eq: reciprocity-beta-inverse}
\end{eqnarray}
which implies that $\beta^{-1}$ is another solution to Eq.~(\ref{eq: characteristic-standard}). Hence, the solutions to the $2lq$-th-order equation~(\ref{eq: characteristic-standard}) can be represented as
\begin{equation}
\left| \beta_{1} \right| \leq \cdots \leq \left| \beta_{lq} \right| \leq 1 \leq | \beta_{lq}^{-1} | \leq \cdots \leq | \beta_{1}^{-1} |.
\end{equation}
Then, using Eq.~(\ref{eq: Yao-Wang-Yokomizo-Murakami}), which is the salient result of the non-Bloch band theory, we have $\left| \beta_{lq} \right| = | \beta_{lq}^{-1} |$, i.e., $\left| \beta_{lq} \right| = 1$. Consequently, bulk eigenstates are delocalized and no skin effects occur.

Notably, Refs.~\cite{Zhang-19, OKSS-20} showed that the skin effects originate from nontrivial topology that cannot be continuously deformed to any Hermitian systems. Consistently, such intrinsic non-Hermitian topology is absent in one-dimensional systems in the orthogonal class; see class $\text{AI}^{\dag}$ in Table~V of Ref.~\cite{KSUS-19}.

\subsection{Symplectic class (class $\text{AII}^{\dag}$)}
	\label{sec: symplectic reciprocity}
	
In the symplectic class, in which Eq.~(\ref{eq: symplectic symmetry}) is respected, we still have Eq.~(\ref{eq: reciprocity-eigenstates}). A crucial distinction is Kramers degeneracy due to $\mathcal{T}\mathcal{T}^{*} = -1$~\cite{Esaki-11, KSUS-19}. Such generic degeneracy is absent in the orthogonal class. In fact, because of $\mathcal{T}^{T} = - \mathcal{T}$, we have
\begin{equation}
\braket{\chi | \mathcal{T} | \chi}^{*}
= \braket{\chi | \mathcal{T}^{T} | \chi}^{*}
= - \braket{\chi | \mathcal{T} | \chi}^{*},
\end{equation}
which leads to $\braket{\chi | \mathcal{T} | \chi}^{*} = 0$. This indicates that $\ket{\phi}$ and $\mathcal{T}\ket{\chi}^{*}$, which belong to the same eigenenergy, are biorthogonal~\cite{Brody-14} to each other and linearly independent of each other. Thus, all the eigenenergies are at least twofold degenerate. 

Similarly to the orthogonal class, we have Eq.~(\ref{eq: reciprocity-beta-inverse}) even in the symplectic class. In terms of $H \left( \beta \right)$, the non-Bloch waves $\ket{\phi_{i}}$ and $\mathcal{T}\ket{\chi_{i}}^{*}$ form a Kramers pair; the former satisfies Eq.~(\ref{eq: H-beta-right}), while the latter satisfies
\begin{equation}
H\,( \beta_{i}^{-1} ) \left( \mathcal{T} \ket{\chi_{i}}^{*} \right) = E \left( \mathcal{T} \ket{\chi_{i}}^{*} \right).
\end{equation} 
Because of this Kramers degeneracy, the characteristic equation has the $4lq$-th order and its solutions are generally represented as
\begin{equation}
\left| \beta_{1} \right| \leq \cdots \leq \left| \beta_{2lq} \right| \leq 1 \leq | \beta_{2lq}^{-1} | \leq \cdots \leq | \beta_{1}^{-1} |.
	\label{eq: beta-symplectic}
\end{equation}
If the standard non-Bloch band theory is applicable, we have $\left| \beta_{2lq} \right| = | \beta_{2lq}^{-1} |$, and hence no skin effects appear in a similar manner to the orthogonal class. However, a reciprocal skin effect is feasible in the symplectic class, as shown in Ref.~\cite{OKSS-20} and the next section. This fact implies modification of the standard non-Bloch band theory, as demonstrated in the following.

\section{Non-Bloch band theory in the symplectic class}
	\label{sec: non-Bloch-symplectic}

We establish the non-Bloch band theory in the symplectic class. In Sec.~\ref{sec: symplectic-Hatano-Nelson}, we begin with exactly solving a symplectic extension of the Hatano-Nelson model~\cite{Hatano-Nelson-96, *Hatano-Nelson-97, OKSS-20} and confirming the skin effect even in the presence of reciprocity. On the basis of this prototypical model, we generally demonstrate our nonstandard non-Bloch band theory in Sec.~\ref{sec: non-Bloch-symplectic; general}. There, Kramers degeneracy in Sec.~\ref{sec: symplectic reciprocity} plays a key role. In Sec.~\ref{sec: sHN-NNN}, we numerically investigate the symplectic Hatano-Nelson model with next-nearest-neighbor hopping to further demonstrate the nonstandard non-Bloch band theory. Because of the reciprocity-protected nature, continuum bands in the symplectic class are fragile against a reciprocity-breaking perturbation, even if it is infinitesimal, as shown in Sec.~\ref{sec: infinitesimal instability}.

\subsection{Symplectic Hatano-Nelson model}
	\label{sec: symplectic-Hatano-Nelson}

The Hatano-Nelson model	~\cite{Hatano-Nelson-96, *Hatano-Nelson-97} is a prototypical non-Hermitian model that exhibits the skin effect, which is given by
\begin{equation}
\hat{H} = \sum_{n} \left[ \left( t+g \right) \hat{c}_{n+1}^{\dag} \hat{c}_{n} + \left( t-g \right) \hat{c}_{n}^{\dag} \hat{c}_{n+1} \right].
\end{equation}
Here, $t \in \mathbb{R}$ is the Hermitian part of the hopping amplitude, and $g \in \mathbb{R}$ describes the asymmetry of the hopping as the degree of non-Hermiticity. The corresponding bulk Hamiltonian defined as Eq.~(\ref{eq: def-bulk-Hamiltonian}) is
\begin{equation}
H \left( \beta \right) = \left( t+g \right) \beta^{-1} + \left( t-g\right) \beta.
\end{equation}
The exact solution to the Hatano-Nelson model is provided, for example, in the Supplemental Material of Ref.~\cite{YM-19}. Under open boundary conditions, we have
\begin{equation}
\left| \beta \right| = \sqrt{\left| \frac{t+g}{t-g} \right|},
\end{equation}
and all the eigenstates are localized at the right (left) edge for $g/t > 0$ ($g/t < 0$). We note in passing that the original works~\cite{Hatano-Nelson-96, *Hatano-Nelson-97} introduced onsite random potential and revealed delocalization transitions even in one dimension due to the interplay between non-Hermiticity and disorder.

Combining a reciprocal pair of the Hatano-Nelson models, we below investigate the following symplectic generalization~\cite{OKSS-20}:
\begin{eqnarray}
\hat{H} &=& \sum_{n} \left[ \hat{c}_{n+1}^{\dag} \left( t - \ii \Delta \sigma_{x} + g\sigma_{z} \right) \hat{c}_{n} \right. \nonumber \\
&&\qquad\qquad\quad \left. + \hat{c}_{n}^{\dag} \left( t + \ii \Delta \sigma_{x} - g\sigma_{z} \right) \hat{c}_{n+1} \right],
\end{eqnarray}
and 
\begin{equation}
H \left( \beta \right) = \left( t - \ii \Delta \sigma_{x} + g\sigma_{z} \right) \beta^{-1} + \left( t + \ii \Delta \sigma_{x} - g\sigma_{z} \right) \beta.
\end{equation}
Here, $\hat{c}_{n}$ ($\hat{c}_{n}^{\dag}$) annihilates (creates) a spinful particle with two components, $\sigma_{i}$'s are Pauli matrices that describe the spin degrees of freedom, and $\Delta$ is the spin-orbit interaction. The bulk Hamiltonian respects 
\begin{equation}
\sigma_{y} H^{T} \left( \beta \right) \sigma_{y}^{-1} = H\,( \beta^{-1} ),
\end{equation}
and indeed belongs to the symplectic class.

\begin{figure}[t]
\centering
\includegraphics[width=86mm]{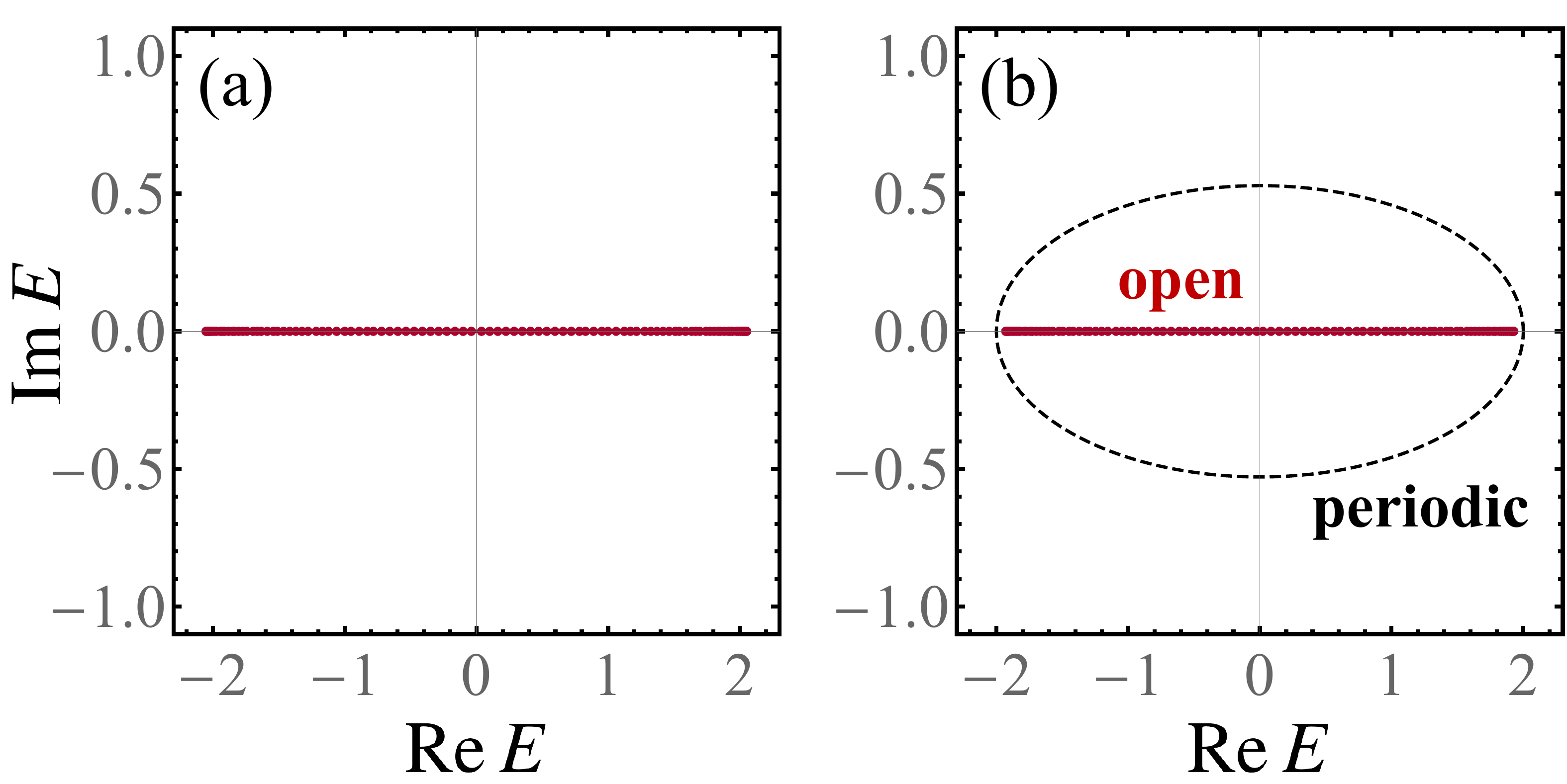} 
\caption{Complex spectra of the symplectic Hatano-Nelson model. The black dashed curves denote the spectra under periodic boundary conditions, and the red dots denote the spectra under open boundary conditions ($L=100$). (a)~The periodic-boundary spectrum and the open-boundary spectrum coincide with each other, and no skin effect occurs ($t=1.0$, $\Delta = 0.3$, $g=0.2$). (b)~The periodic-boundary spectrum forms a loop in the complex plane, but the open-boundary spectrum lies on the real axis, which is a signature of the skin effect ($t=1.0$, $\Delta = 0.3$, $g=0.4$).}
	\label{fig: sHN-spectra}
\end{figure}

Under periodic boundary conditions, $\beta$ satisfies $\left| \beta \right| = 1$ and hence is given by $\beta := e^{\ii k}$ with real wavenumbers $k \in \left[ 0, 2\pi \right]$. Then, the Bloch Hamiltonian is 
\begin{equation}
H \left( k \right) = 2t \cos k - 2 \left( \Delta \sigma_{x} + \ii g \sigma_{z} \right) \sin k.
\end{equation}
The spectrum of $H \left( k \right)$ is given as
\begin{equation}
E \left( k \right) = 2t \cos k \pm 2 \ii \sqrt{g^{2} - \Delta^{2}} \sin k,
\end{equation}
which is entirely real for $\left|g \right| \leq \left| \Delta \right|$ and forms a loop in the complex plane for $\left|g \right| > \left| \Delta \right|$ (Fig.~\ref{fig: sHN-spectra}).

Under open boundary conditions, let $E \in \mathbb{C}$ be an eigenenergy and $\ket{\phi} = \sum_{n=1}^{L} \sum_{s \in \{ \uparrow, \downarrow \}} \phi_{n, s} \ket{n} \ket{s}$ be the corresponding right eigenstate. The Schr\"odinger equation in real space reads
\begin{eqnarray}
&&\left( t - \ii \Delta \sigma_{x} + g\sigma_{z} \right) \begin{pmatrix}
\phi_{n-1, \uparrow} \\ \phi_{n-1, \downarrow}
\end{pmatrix} \nonumber \\
&&\quad+ \left( t + \ii \Delta \sigma_{x} - g\sigma_{z} \right) \begin{pmatrix}
\phi_{n+1, \uparrow} \\ \phi_{n+1, \downarrow}
\end{pmatrix}
= E \begin{pmatrix}
\phi_{n, \uparrow} \\ \phi_{n, \downarrow}
\end{pmatrix}
	\label{eq: sHN-Sch-bulk}
\end{eqnarray}
in the bulk ($n=2, 3, \cdots, L-1$), and
\begin{eqnarray}
\left( t + \ii \Delta \sigma_{x} - g\sigma_{z} \right) \begin{pmatrix}
\phi_{2, \uparrow} \\ \phi_{2, \downarrow}
\end{pmatrix} &=& E \begin{pmatrix}
\phi_{1, \uparrow} \\ \phi_{1, \downarrow}
\end{pmatrix}, 
	\label{eq: sHN-Sch-edge1} \\
\left( t - \ii \Delta \sigma_{x} + g\sigma_{z} \right) \begin{pmatrix}
\phi_{L-1, \uparrow} \\ \phi_{L-1, \downarrow}
\end{pmatrix} &=& E \begin{pmatrix}
\phi_{L, \uparrow} \\ \phi_{L, \downarrow}
\end{pmatrix}
	\label{eq: sHN-Sch-edge2}
\end{eqnarray}
at the edges. Defining $\phi_{0, s}$ and $\phi_{L+1, s}$ by the bulk equation~(\ref{eq: sHN-Sch-bulk}), the boundary equations~(\ref{eq: sHN-Sch-edge1}) and (\ref{eq: sHN-Sch-edge2}) reduce to
\begin{eqnarray}
\begin{pmatrix}
\phi_{0, \uparrow} \\ \phi_{0, \downarrow}
\end{pmatrix} = \begin{pmatrix}
\phi_{L+1, \uparrow} \\ \phi_{L+1, \downarrow}
\end{pmatrix} = 0.
	\label{eq: sHN-Sch-edge34}
\end{eqnarray}

Suppose a fundamental solution is given as $\phi_{n, s} \propto \beta^{n} \phi_{s}$. From the bulk equation~(\ref{eq: sHN-Sch-bulk}), we have
\begin{equation}
\left[ H \left( \beta \right) - E \right] \begin{pmatrix}
\phi_{\uparrow} \\ \phi_{\downarrow}
\end{pmatrix} = 0.
	\label{eq: sHN-bulk-eigenequation}
\end{equation}
To have a nontrivial solution $\left( \phi_{\uparrow}~\phi_{\downarrow} \right)^{T} \neq 0$, the coefficient matrix $H \left( \beta \right) - E$ should not be invertible, leading to the characteristic equation
\begin{equation}
\det \left[ H \left( \beta \right) - E \right] = 0,
\end{equation}
i.e.,
\begin{equation}
E = t \left( \beta + \beta^{-1} \right) \pm \sqrt{g^{2}-\Delta^{2}} \left( \beta- \beta^{-1} \right).
\end{equation}
This is a quartic equation in terms of $\beta$ for given $E$ and decomposes into a pair of quadratic equations
\begin{eqnarray}
\left( t + \sqrt{g^{2}-\Delta^{2}} \right) \beta^{2} - E \beta + t - \sqrt{g^{2}-\Delta^{2}} &=& 0,\quad	\label{eq: quadratic-1}\\
\left( t - \sqrt{g^{2}-\Delta^{2}} \right) \beta^{2} - E \beta + t + \sqrt{g^{2}-\Delta^{2}} &=& 0. \label{eq: quadratic-2}
\end{eqnarray}
Remarkably, when $\beta$ satisfies this characteristic equation, $\beta^{-1}$ also satisfies it; in particular, when $\beta$ satisfies Eq.~(\ref{eq: quadratic-1}), $\beta^{-1}$ satisfies Eq.~(\ref{eq: quadratic-2}), and vice versa. This is a direct consequence of reciprocity, as discussed in Sec.~\ref{sec: reciprocity}. Furthermore, a fundamental solution with $\beta$ and another fundamental solution with $\beta^{-1}$ are linearly independent of each other and form a Kramers pair. Now, we define the solutions to Eq.~(\ref{eq: quadratic-1}) as $\beta_{1}$ and $\beta_{2}$ ($\left| \beta_{1} \right| \leq \left| \beta_{2} \right|$), which satisfy
\begin{equation}
\beta_{1} \beta_{2} = \frac{t-\sqrt{g^{2} - \Delta^{2}}}{t+\sqrt{g^{2} - \Delta^{2}}}.
	\label{eq: sHN-beta-product}
\end{equation}
The solutions to Eq.~(\ref{eq: quadratic-2}) are given as $\beta_{1}^{-1}$ and $\beta_{2}^{-1}$. Since the solutions $\beta_{1}, \beta_{2}, \beta_{1}^{-1}, \beta_{2}^{-1}$ to the characteristic equation are defined to respect $| \beta_{1} | \leq | \beta_{2} |$, the standard non-Bloch band theory~\cite{YW-18-SSH, YM-19} predicts $\left| \beta_{2} \right| = | \beta_{2}^{-1} |$ for continuum bands. However, this is not the case in the symplectic class; we have $\left| \beta_{1} \right| = \left| \beta_{2} \right|$, as shown below.

Now, the eigenstate $\ket{\phi} = \sum_{n=1}^{L} \sum_{s \in \{ \uparrow, \downarrow \}} \phi_{n, s} \ket{n} \ket{s}$ can be obtained as a linear combination of the above fundamental solutions:
\begin{eqnarray}
\begin{pmatrix}
\phi_{n, \uparrow} \\ \phi_{n, \downarrow}
\end{pmatrix} 
&=& \beta_{1}^{n} \begin{pmatrix}
\phi_{\uparrow}^{(1+)} \\ \phi_{\downarrow}^{(1+)}
\end{pmatrix} + \beta_{2}^{n} \begin{pmatrix}
\phi_{\uparrow}^{(2+)} \\ \phi_{\downarrow}^{(2+)}
\end{pmatrix} \nonumber \\
&&\quad+ \beta_{1}^{-n} \begin{pmatrix}
\phi_{\uparrow}^{(1-)} \\ \phi_{\downarrow}^{(1-)}
\end{pmatrix} + \beta_{2}^{-n} \begin{pmatrix}
\phi_{\uparrow}^{(2-)} \\ \phi_{\downarrow}^{(2-)}
\end{pmatrix}
	\label{eq: sHN-generic-eigenstate}
\end{eqnarray}
for $n = 1, 2, \cdots, L$. Here, since $( \phi_{\uparrow}^{(i\pm)}~\phi_{\downarrow}^{(i\pm)} )^{T}$ satisfies Eq.~(\ref{eq: sHN-bulk-eigenequation}), we have
\begin{equation}
\begin{pmatrix}
\pm \sqrt{g^{2}-\Delta^{2}} + g & -\ii \Delta \\
- \ii \Delta & \pm \sqrt{g^{2}-\Delta^{2}} - g
\end{pmatrix} \begin{pmatrix}
\phi_{\uparrow}^{(i\pm)} \\ \phi_{\downarrow}^{(i\pm)}
\end{pmatrix} = 0.
\end{equation}
Remarkably, $( \phi_{\uparrow}^{(i\pm)}~\phi_{\downarrow}^{(i\pm)} )^{T}$ does not depend on $i$. Hence, Eq.~(\ref{eq: sHN-generic-eigenstate}) is further simplified to
\begin{eqnarray}
\begin{pmatrix}
\phi_{n, \uparrow} \\ \phi_{n, \downarrow}
\end{pmatrix} 
&=& ( \beta_{1}^{n} \bar{\phi}_{1+} + \beta_{2}^{n} \bar{\phi}_{2+} ) \begin{pmatrix}
1 \\ c_{+}
\end{pmatrix} \nonumber \\
&&\qquad + ( \beta_{1}^{-n} \bar{\phi}_{1-} + \beta_{2}^{-n} \bar{\phi}_{2-} ) \begin{pmatrix}
1 \\ c_{-}
\end{pmatrix},
\end{eqnarray}
with some constants $\bar{\phi}_{1\pm}, \bar{\phi}_{2\pm} \in \mathbb{C}$ and $c_{\pm} := -\ii\,( \pm\sqrt{g^{2}-\Delta^{2}} + g )/\Delta$. Then, the boundary condition~(\ref{eq: sHN-Sch-edge34}) reduces to
\begin{eqnarray}
&&( \bar{\phi}_{1+} + \bar{\phi}_{2+} ) \begin{pmatrix}
1 \\ c_{+}
\end{pmatrix}
+ ( \bar{\phi}_{1-} + \bar{\phi}_{2-} ) \begin{pmatrix}
1 \\ c_{-}
\end{pmatrix} = 0,\\
&&( \beta_{1}^{L+1} \bar{\phi}_{1+} + \beta_{2}^{L+1} \bar{\phi}_{2+} ) \begin{pmatrix}
1 \\ c_{+}
\end{pmatrix} \nonumber \\
&&\qquad+\,( \beta_{1}^{- \left( L+1 \right)} \bar{\phi}_{1-} + \beta_{2}^{- \left( L+1 \right)} \bar{\phi}_{2-} ) \begin{pmatrix}
1 \\ c_{-}
\end{pmatrix} = 0.
\end{eqnarray}
The vectors $( 1~c_{+} )^{T}$ and $( 1~c_{-} )^{T}$ form a Kramers pair and are linearly independent of each other. In particular, they are biorthogonal to each other~\cite{Brody-14}, i.e., the left counterpart of $( 1~c_{+} )^{T}$ is orthogonal to $( 1~c_{-} )^{T}$. As a result, we have
\begin{equation}
\bar{\phi}_{1\pm} + \bar{\phi}_{2\pm} = \beta_{1}^{\pm \left( L+1 \right)} \bar{\phi}_{1\pm} + \beta_{2}^{\pm \left( L+1 \right)} \bar{\phi}_{2\pm} = 0,
	\label{eq: sHN-bc}
\end{equation}
leading to
\begin{equation}
\beta_{1}^{L+1} = \beta_{2}^{L+1}
\end{equation}
for a nontrivial solution $\left( \bar{\phi}_{1\pm}, \bar{\phi}_{2\pm} \right) \neq 0$. This equation means that the absolute values of $\beta_{1}$ and $\beta_{2}$ coincide with each other and are given, from Eq.~(\ref{eq: sHN-beta-product}), as
\begin{equation}
\left| \beta_{1} \right| = \left| \beta_{2} \right| =
\sqrt{\left| \frac{t-\sqrt{g^{2} - \Delta^{2}}}{t+\sqrt{g^{2} - \Delta^{2}}}\right|}.
	\label{eq: sHN-amplitude}
\end{equation}
The relative phase between $\beta_{1}$ and $\beta_{2}$ can be different, resulting in the formation of continuum bands.

Equation~(\ref{eq: sHN-amplitude}) provides the localization length of eigenstates and the criteria of the skin effect. For $\left| g \right| \leq \left| \Delta \right|$, we have $\left| \beta_{1} \right| = \left| \beta_{2} \right| = 1$ and hence eigenstates are delocalized. For $\left| g \right| > \left| \Delta \right|$, on the other hand, we have $\left| \beta_{1} \right| = \left| \beta_{2} \right| \neq 1$ and hence eigenstates are localized at the edges. In contrast to the conventional skin effect, skin modes appear at both edges; when an eigenstate is localized at one edge, the Kramers partner is localized at the other edge. The numerical calculations shown in Fig.~\ref{fig: sHN-spectra} confirm this result. For $\left| g \right| > \left| \Delta \right|$, the spectrum under periodic boundary conditions forms a loop in the complex plane, but the spectrum under open boundary conditions lies on the real axis, which is a signature of the non-Hermitian skin effect.

In the above calculations, an important distinction from the standard case is the equivalence between $( \phi_{\uparrow}^{(1\pm)}~\phi_{\downarrow}^{(1\pm)} )^{T}$ and $( \phi_{\uparrow}^{(2\pm)}~\phi_{\downarrow}^{(2\pm)} )^{T}$. In fact, if they were linearly independent, we would have $|\beta_{2}| = |\beta_{2}^{-1}|$ instead of $|\beta_{1}| = |\beta_{2}|$ in a similar manner to the standard case~\cite{YW-18-SSH, YM-19}. However, symplectic reciprocity makes $( \phi_{\uparrow}^{(1\pm)}~\phi_{\downarrow}^{(1\pm)} )^{T}$ and $( \phi_{\uparrow}^{(2\pm)}~\phi_{\downarrow}^{(2\pm)} )^{T}$ linearly dependent on each other and changes the condition for continuum bands, as demonstrated below.

\subsection{General condition}
	\label{sec: non-Bloch-symplectic; general}
	
Now, we demonstrate the non-Bloch band theory in the symplectic class in a general manner. We consider a generic non-Hermitian Hamiltonian described by 
\begin{equation}
\hat{H} = \sum_{n} \sum_{j=-l}^{l} \sum_{\mu, \nu=1}^{q} \sum_{s, t \in \{ \uparrow, \downarrow \}} H_{j, \mu\nu, st} \hat{c}_{n+j, \mu, s}^{\dag} \hat{c}_{n, \nu, t}.
\end{equation}
In comparison with the standard class discussed in Sec.~\ref{sec: standard non-Bloch}, the indices $s, t \in \{ \uparrow, \downarrow \}$ are added to Eq.~(\ref{eq: Hamiltonian-orthogonal}) to account for the internal degrees of freedom arising from Eq.~(\ref{eq: symplectic symmetry}). A prime example of such internal degrees of freedom is the spin degrees of freedom. Correspondingly, the eigenstates are given by a linear combination of fundamental solutions
\begin{equation}
\sum_{n=1}^{L} \ket{n} \left( \beta^{n}_{i} \ket{\phi_{i+}} + \beta^{-n}_{i} \ket{\phi_{i-}} \right),
\end{equation}
where $\ket{\phi_{i\pm}}$ can be expanded as
\begin{equation}
\ket{\phi_{i\pm}} := \sum_{\mu=1}^{q} \sum_{s \in \{ \uparrow, \downarrow \}} \phi_{\mu s}^{(i\pm)} \ket{\mu} \ket{s},
	\label{eq: generic-eigenstates-symplectic-components}
\end{equation}
and is a right eigenstate of $H\,( \beta_{i}^{\pm} )$:
\begin{equation}
H\,( \beta_{i}^{\pm} ) \ket{\phi_{i\pm}} = E \ket{\phi_{i\pm}}.
	\label{eq: Schroedinger-bulk-symplectic}
\end{equation}
The corresponding left eigenstate $\ket{\chi_{i\pm}}$ of $H\,( \beta_{i}^{\pm} )$ is defined by
\begin{equation}
H^{\dag} \,( \beta_{i}^{\pm} ) \ket{\chi_{i\pm}} = E^{*} \ket{\chi_{i\pm}}.
\end{equation}
Here, $\left\{ \pm \right\}$ is equivalent to $\left\{ \uparrow, \downarrow \right\}$ in the absence of perturbations that mix $\uparrow$ and $\downarrow$, including spin-orbit interaction; but this is not necessarily true, in general. In addition, $\beta_{i}$'s and $\beta_{i}^{-1}$'s ($i = 1, 2, \cdots, 2lq$) are the solutions to the characteristic equation $\det \left[ H \left( \beta \right) - E \right] = 0$. Without loss of generality, they can be chosen so that Eq.~(\ref{eq: beta-symplectic}) will be satisfied. Importantly, as described in Sec.~\ref{sec: symplectic reciprocity}, $\ket{\phi_{i+}}$ and $\ket{\phi_{i-}}$ are biorthogonal to each other and form a Kramers pair. Specifically, we have
\begin{equation}
\ket{\phi_{i-}} = \mathcal{T} \ket{\chi_{i+}}^{*},\quad
\ket{\phi_{i+}} = -\mathcal{T} \ket{\chi_{i-}}^{*}
	\label{eq: Kramers-normalization}
\end{equation}
under the appropriate choice of the gauges. Generally, the left eigenstates $\ket{\chi_{i\pm}}$'s are determined when the right eigenstates $\ket{\phi_{i\pm}}$'s are given, except for the arbitrariness of normalization~\cite{Brody-14}. In this respect, Eq.~(\ref{eq: Kramers-normalization}) provides normalization conditions of $\ket{\chi_{i\pm}}$'s. From these fundamental solutions, the right eigenstate $\ket{\phi}$ and the left eigenstate $\ket{\chi}$ in real space can be given as 
\begin{eqnarray}
\ket{\phi} &=& \sum^{2lq}_{i=1} \sum_{n=1}^{L} \ket{n} \left( \beta^{n}_{i} \ket{\phi_{i+}} + \beta^{-n}_{i} \ket{\phi_{i-}} \right), \label{eq: generic-eigenstates-symplectic} \\
\ket{\chi} &=& \sum^{2lq}_{i=1} \sum_{n=1}^{L} \ket{n} \left( \left( \beta_{i}^{*} \right)^{-n} \ket{\chi_{i+}} + \left( \beta_{i}^{*} \right)^{n} \ket{\chi_{i-}} \right)
\end{eqnarray}
under the appropriate choice of the gauges and normalization. In addition, we have from Eq.~(\ref{eq: Kramers-normalization})
\begin{equation}
\mathcal{T} \ket{\chi}^{*} = \sum^{2lq}_{i=1} \sum_{n=1}^{L} \ket{n} \left( - \beta_{i}^{n} \ket{\phi_{i+}} + \beta_{i}^{-n} \ket{\phi_{i-}} \right).
	\label{eq: generic-eigenstates-symplectic-Kramers-partner}
\end{equation}
which is the Kramers partner of $\ket{\phi}$ satisfying $\braket{\chi | \mathcal{T} | \chi}^{*} = 0$.

Generic eigenstates $\ket{\phi}$ and $\mathcal{T} \ket{\chi}^{*}$ in Eqs.~(\ref{eq: generic-eigenstates-symplectic}) and (\ref{eq: generic-eigenstates-symplectic-Kramers-partner}) include $2lq \times 2 \times q \times 2$ unknown variables $\phi_{\mu s}^{(i\pm)}$ ($i = 1, 2, \cdots, 2lq$; $\mu = 1, 2, \cdots, q$; $s= \uparrow, \downarrow$) in Eq.~(\ref{eq: generic-eigenstates-symplectic-components}). They reduce to $2lq \times 2$ unknown variables, for example, $\phi_{1\uparrow}^{(i\pm)}$, because of the Schr\"odinger equation~(\ref{eq: Schroedinger-bulk-symplectic}) for the bulk Hamiltonian $H \left( \beta \right)$. Here, the rank of $H \left( \beta \right)$ is assumed appropriately in a similar manner to Ref.~\cite{YM-19}.

The $2lq \times 2$ unknown variables $\bar{\phi}_{i\pm} := \phi_{1\uparrow}^{(i\pm)}$ ($i = 1, \cdots, 2lq$) are determined by boundary conditions. In general, the boundary conditions are given by details about the $lq$ sites around each end. Hence, the boundary conditions for $\ket{\phi}$ can be represented by
\begin{eqnarray}
&&\sum_{i=1}^{2lq} \left( f_{j} \left( \beta_{i} \right) \bar{\phi}_{i+} + f_{j}\,( \beta_{i}^{-1} )\,\bar{\phi}_{i-} \right) 
= 0, \label{eq: general-symplectic-bc1} \\
&&\sum_{i=1}^{2lq} \left( \beta_{i}^{L} g_{j} \left( \beta_{i} \right) \bar{\phi}_{i+} + \beta_{i}^{- L} g_{j}\,( \beta_{i}^{-1} )\,\bar{\phi}_{i-} \right) 
= 0,\label{eq: general-symplectic-bc2}
\end{eqnarray}
where $f_{j} \left( \beta_{i} \right)$ and $g_{j}  \left( \beta_{i} \right)$ ($j = 1, 2, \cdots, 2lq$) are functions of $\beta_{i}$ that do not depend on $L$. Importantly, the boundary conditions for the Kramers partner $\mathcal{T}\ket{\chi}^{*}$ are independent of those for $\ket{\phi}$, and should also be respected:
\begin{eqnarray}
&&\sum_{i=1}^{2lq} \left( -f_{j} \left( \beta_{i} \right) \bar{\phi}_{i+} + f_{j}\,( \beta_{i}^{-1} )\,\bar{\phi}_{i-} \right) 
= 0, \label{eq: general-symplectic-bc3} \\
&&\sum_{i=1}^{2lq} \left( -\beta_{i}^{L} g_{j} \left( \beta_{i} \right) \bar{\phi}_{i+} + \beta_{i}^{- L} g_{j}\,( \beta_{i}^{-1} )\,\bar{\phi}_{i-} \right) 
= 0. \label{eq: general-symplectic-bc4}
\end{eqnarray}
For example, when the eigenstates vanish at $n=0$ and $n=L+1$ in a similar manner to the example in Sec.~\ref{sec: symplectic-Hatano-Nelson}, we have $\sum_{i=1}^{2lq} \left( \ket{\phi_{i+}} + \ket{\phi_{i-}} \right) = \sum_{i=1}^{2lq}\,( \beta_{i}^{L+1} \ket{\phi_{i+}} + \beta_{i}^{- \left( L+1 \right)} \ket{\phi_{i-}} ) = 0$ for $\ket{\phi}$, and $f_{j} \left( \beta_{i} \right)$ and $g_{j}  \left( \beta_{i} \right)$ are given by $\ket{\phi_{i\pm}}$. Indeed, the boundary condition~(\ref{eq: sHN-bc}) is described by these equations.

Whereas we have $4lq$ unknown variables $\bar{\phi}_{i\pm}$, the boundary conditions~(\ref{eq: general-symplectic-bc1})-(\ref{eq: general-symplectic-bc4}) provide $8lq$ linear equations. This implies that Eqs.~(\ref{eq: general-symplectic-bc1})-(\ref{eq: general-symplectic-bc4}) are not linearly independent of each other because of some constraints on $f_{j} \left( \beta_{i} \right)$ and $g_{j}  \left( \beta_{i} \right)$. To have such constraints, we notice from Eqs.~(\ref{eq: general-symplectic-bc1}) and (\ref{eq: general-symplectic-bc3}) 
\begin{equation}
\sum_{i=1}^{2lq} f_{j} \left( \beta_{i} \right) \bar{\phi}_{i+} 
= \sum_{i=1}^{2lq} f_{j}\,( \beta_{i}^{-1} )\,\bar{\phi}_{i-} 
= 0.
\end{equation}
In matrix representation, we have
\begin{equation}
\begin{pmatrix}
f_{1}\,( \beta_{1}^{\pm} ) & \cdots & f_{1}\,( \beta_{2lq}^{\pm} ) \\
\vdots & \ddots & \vdots \\
f_{2lq}\,( \beta_{1}^{\pm} ) & \cdots & f_{2lq}\,( \beta_{2lq}^{\pm} )
\end{pmatrix} \begin{pmatrix}
\bar{\phi}_{1\pm} \\ \vdots \\ \bar{\phi}_{2lq\pm}
\end{pmatrix} = 0.
\end{equation}
To have a nontrivial solution, the $2lq \times 2lq$ coefficient matrix $F_{\pm}$ should not be invertible, i.e., 
\begin{equation}
\det F_{\pm}
:= \det \begin{pmatrix}
f_{1}\,( \beta_{1}^{\pm} ) & \cdots & f_{1}\,( \beta_{2lq}^{\pm} ) \\
\vdots & \ddots & \vdots \\
f_{2lq}\,( \beta_{1}^{\pm} ) & \cdots & f_{2lq}\,( \beta_{2lq}^{\pm} )
\end{pmatrix}
= 0.
	\label{eq: det F}
\end{equation}
Consistently, these constraints are respected in the specific example in Sec.~\ref{sec: symplectic-Hatano-Nelson} since $( \phi_{\uparrow}^{(1\pm)}~\phi_{\downarrow}^{(1\pm)} )^{T}$ and $( \phi_{\uparrow}^{(2\pm)}~\phi_{\downarrow}^{(2\pm)} )^{T}$ are linearly dependent on each other. The combination of Eqs.~(\ref{eq: general-symplectic-bc2}) and (\ref{eq: general-symplectic-bc4}) yields similar constraints on $g_{j}  \left( \beta_{i} \right)$.

Then, from Eqs.~(\ref{eq: general-symplectic-bc1}) and (\ref{eq: general-symplectic-bc2}), we have
\begin{widetext}
\begin{equation}
\begin{pmatrix}
f_{1} \left( \beta_{1} \right) & \cdots & f_{1}\,( \beta_{2lq} ) & f_{1}\,( \beta_{1}^{-1} ) & \cdots & f_{1}\,( \beta_{2lq}^{-1} ) \\
\vdots & \ddots & \vdots & \vdots & \ddots & \vdots \\
f_{2lq} \left( \beta_{1} \right) & \cdots & f_{2lq} \left( \beta_{2lq} \right) & f_{2lq}\,( \beta_{1}^{-1} ) & \cdots & f_{2lq}\,( \beta_{2lq}^{-1} ) \\
\beta_{1}^{L} g_{1}\,( \beta_{1} ) & \cdots & \beta_{2lq}^{L} g_{1}\,( \beta_{2lq} ) & \beta_{1}^{-L} g_{1}\,( \beta_{1}^{-1} ) & \cdots & \beta_{2lq}^{-L} g_{1}\,( \beta_{2lq}^{-1} ) \\
\vdots & \ddots & \vdots & \vdots & \ddots & \vdots \\
\beta_{1}^{L} g_{2lq}\,( \beta_{1} ) & \cdots & \beta_{2lq}^{L} g_{2lq}\,( \beta_{2lq} ) & \beta_{1}^{-L} g_{2lq}\,( \beta_{1}^{-1} ) & \cdots & \beta_{2lq}^{-L} g_{2lq}\,( \beta_{2lq}^{-1} )
\end{pmatrix} \begin{pmatrix}
\bar{\phi}_{1+} \\ \vdots \\ \bar{\phi}_{2lq+} \\ \bar{\phi}_{1-} \\ \vdots \\ \bar{\phi}_{2lq-}
\end{pmatrix} = 0.
\end{equation}
To have a nontrivial solution, the $4lq \times 4lq$ coefficient matrix should not be invertible, i.e., 
\begin{equation}
\det \begin{pmatrix}
f_{1} \left( \beta_{1} \right) & \cdots & f_{1}\,( \beta_{2lq} ) & f_{1}\,( \beta_{1}^{-1} ) & \cdots & f_{1}\,( \beta_{2lq}^{-1} ) \\
\vdots & \ddots & \vdots & \vdots & \ddots & \vdots \\
f_{2lq} \left( \beta_{1} \right) & \cdots & f_{2lq} \left( \beta_{2lq} \right) & f_{2lq}\,( \beta_{1}^{-1} ) & \cdots & f_{2lq}\,( \beta_{2lq}^{-1} ) \\
\beta_{1}^{L} g_{1}\,( \beta_{1} ) & \cdots & \beta_{2lq}^{L} g_{1}\,( \beta_{2lq} ) & \beta_{1}^{-L} g_{1}\,( \beta_{1}^{-1} ) & \cdots & \beta_{2lq}^{-L} g_{1}\,( \beta_{2lq}^{-1} ) \\
\vdots & \ddots & \vdots & \vdots & \ddots & \vdots \\
\beta_{1}^{L} g_{2lq}\,( \beta_{1} ) & \cdots & \beta_{2lq}^{L} g_{2lq}\,( \beta_{2lq} ) & \beta_{1}^{-L} g_{2lq}\,( \beta_{1}^{-1} ) & \cdots & \beta_{2lq}^{-L} g_{2lq}\,( \beta_{2lq}^{-1} )
\end{pmatrix} = 0.
	\label{eq: det-symplectic}
\end{equation}
\end{widetext}
The determinant on the left-hand side is a $2lq$-th-order polynomial in terms of $\beta_{1}^{L}, \cdots, \beta_{2lq}^{L}, \beta_{2lq}^{-L}, \cdots, \beta_{1}^{-L}$. Because of Eq.~(\ref{eq: beta-symplectic}), its leading-order term includes $( \beta_{2lq}^{-1} \beta_{2lq-1}^{-1} \cdots \beta_{1}^{-1} )^{L}$ and the next-to-leading-order term includes $( \beta_{2lq} \beta_{2lq-1}^{-1} \beta_{2lq-2}^{-1}\cdots \beta_{1}^{-1} )^{L}$, in general. To satisfy Eq.~(\ref{eq: det-symplectic}) for $L \to \infty$, the absolute values of these terms need to coincide with each other, which leads to the condition~(\ref{eq: Yao-Wang-Yokomizo-Murakami}) for the standard case~\cite{YW-18-SSH, YM-19}. However, this is not the case in the symplectic class. In fact, the term including $( \beta_{2lq}^{-1} \beta_{2lq-1}^{-1} \cdots \beta_{1}^{-1} )^{L}$ does not appear since it is proportional to $\det F_{+}$, which vanishes as shown in Eq.~(\ref{eq: det F}). As a result, the leading-order term includes $( \beta_{2lq} \beta_{2lq-1}^{-1} \beta_{2lq-2}^{-1} \cdots \beta_{1}^{-1} )^{L}$ and the next-to-leading-order term includes $( \beta_{2lq-2}^{-1} \beta_{2lq-3}^{-1} \cdots \beta_{1}^{-1} )^{L}$, both of which should be comparable to each other for $L \to \infty$. Therefore, it is necessary to have 
\begin{equation}
| \beta_{2lq} \beta_{2lq-1}^{-1} \beta_{2lq-2}^{-1} \cdots \beta_{1}^{-1} | = | \beta_{2lq-2}^{-1} \beta_{2lq-3}^{-1} \cdots \beta_{1}^{-1} |,
\end{equation}
leading to $|\beta_{2lq-1}| = |\beta_{2lq}|$, i.e., Eq.~(\ref{eq: Yao-Wang-Yokomizo-Murakami-symplectic}) with $M := lq$.

\subsection{Symplectic Hatano-Nelson model with next-nearest-neighbor hopping}
	\label{sec: sHN-NNN}

\begin{figure}[b]
\centering
\includegraphics[width=86mm]{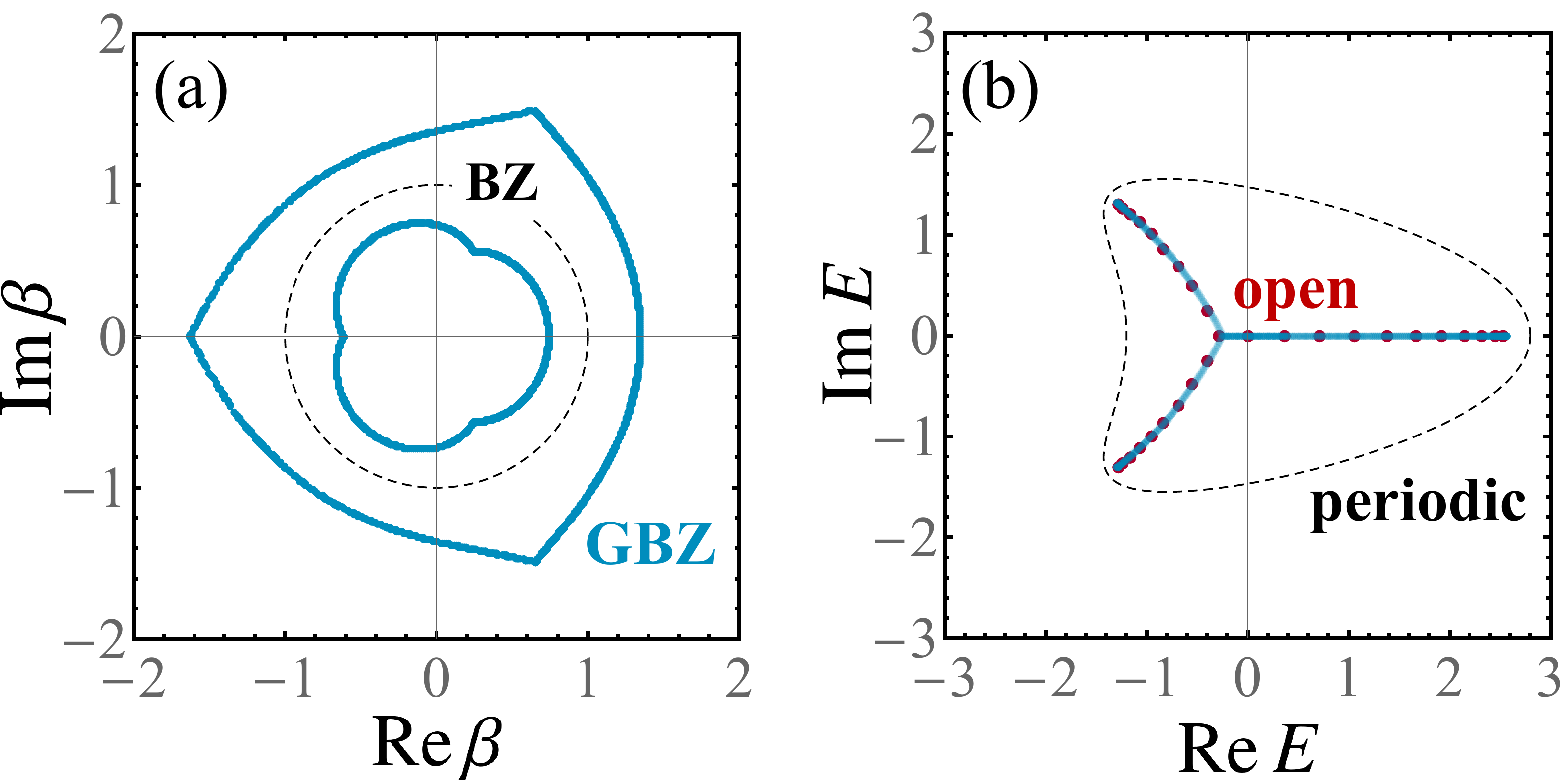} 
\caption{Symplectic Hatano-Nelson model with next-nearest-neighbor hopping ($t=1.0$, $\Delta = 0.2$, $g=0.8$, $t'=0.4$). (a)~Brillouin zone (BZ) and generalized Brillouin zone (GBZ). The system with periodic boundaries is described by the BZ (black dashed loop), which forms the unit circle in the complex $\beta$ plane. By contrast, the system with open boundaries is described by the GBZ (blue solid loops), which is determined by the non-Bloch band theory in the symplectic class. (b)~Complex spectra. The open-boundary spectrum ($L=30$, red dots) coincides with the non-Bloch bands determined by the GBZ (blue solid curves), which is different from the periodic-boundary spectrum (black dashed loop).}
	\label{fig: sHN-NNN}
\end{figure}

To further verify the nonstandard non-Bloch band theory, we consider the symplectic Hatano-Nelson model with next-nearest-neighbor hopping:
\begin{eqnarray}
\hat{H} &=& \sum_{n} \left[ \hat{c}_{n+1}^{\dag} \left( t - \ii \Delta \sigma_{x} + g\sigma_{z} \right) \hat{c}_{n} \right. \nonumber \\
&&\qquad\quad + \hat{c}_{n}^{\dag} \left( t + \ii \Delta \sigma_{x} - g\sigma_{z} \right) \hat{c}_{n+1} \nonumber \\
&&\qquad\qquad\qquad \left. + t'\,( \hat{c}_{n+2}^{\dag} \hat{c}_{n} + \hat{c}_{n}^{\dag} \hat{c}_{n+2} ) \right],
\end{eqnarray}
where $t'$ denotes the amplitude of the next-nearest-neighbor hopping. The bulk Hamiltonian reads
\begin{eqnarray}
H \left( \beta \right) &=& \left( t - \ii \Delta \sigma_{x} + g\sigma_{z} \right) \beta^{-1} + \left( t + \ii \Delta \sigma_{x} - g\sigma_{z} \right) \beta \nonumber \\
&&\qquad\qquad\qquad\qquad\qquad +t' \left( \beta^{2} + \beta^{-2} \right),
\end{eqnarray}
and the energy dispersion is given by the characteristic equation $\det \left[ H \left( \beta \right) - E \right] = 0$, i.e., 
\begin{equation}
E = t \left( \beta + \beta^{-1} \right) + t' \left( \beta^{2} + \beta^{-2} \right) \pm \sqrt{g^{2}-\Delta^{2}} \left( \beta- \beta^{-1} \right).
	\label{eq: characteristic equation - NNN}
\end{equation}
The characteristic equation is eighth order and decomposes into a pair of quartic equations in terms of $\beta$, while it reduces to a pair of quadratic equations for $t'=0$ as investigated in Sec.~\ref{sec: symplectic-Hatano-Nelson}. Consequently, the analytical solutions are not simple.

We numerically investigate the symplectic Hatano-Nelson model with the next-nearest-neighbor hopping in a similar manner to Ref.~\cite{YM-19} and confirm that it is indeed described by the nonstandard non-Bloch band theory. Figure~\ref{fig: sHN-NNN}\,(a) shows the Brillouin zone and the generalized Brillouin zone of this model. The system with periodic boundaries is described by the Brillouin zone, which forms the unit circle $\left| \beta \right| = 1$ in the complex $\beta$ plane. By contrast, the system with open boundaries is described by the generalized Brillouin zone, which is determined by Eq.~(\ref{eq: Yao-Wang-Yokomizo-Murakami-symplectic}) of the non-Bloch band theory in the symplectic class. In this model, Eq.~(\ref{eq: Yao-Wang-Yokomizo-Murakami-symplectic}) means $\left| \beta_{3} \right| = \left| \beta_{4} \right|$, where the eight solutions to Eq.~(\ref{eq: characteristic equation - NNN}) for given $E \in \mathbb{C}$ are denoted as $\beta_{1}, \beta_{2}, \beta_{3}, \beta_{4}; \beta_{4}^{-1}, \beta_{3}^{-1}, \beta_{2}^{-1}, \beta_{1}^{-1}$ with $\left| \beta_{1} \right| \leq \left| \beta_{2} \right| \leq \left| \beta_{3} \right| \leq \left| \beta_{4} \right|$. In the presence of non-Hermiticity, the generalized Brillouin zone does not necessarily form the unit circle, which is a direct manifestation of the non-Hermitian skin effect. In contrast to the standard case, the generalized Brillouin zone generally consists of a pair of loops, one of which is inside the unit circle and the other of which is outside the unit circle. This is a consequence of reciprocity, and each loop describes the localized modes at the right or left edge. Figure~\ref{fig: sHN-NNN}\,(b) shows the complex spectra of this model. The open-boundary spectrum cannot be described by the Bloch bands determined by the Brillouin zone, which is another signature of the skin effect. However, it is in complete agreement with the non-Bloch bands determined by the generalized Brillouin zone. It is also notable that both generalized Brillouin zone and complex spectra are symmetric about the real axis, which originates from time-reversal symmetry $\sigma_{z} H^{*} \left( \beta \right) \sigma_{z}^{-1} = H \left( \beta^{*} \right)$.

\subsection{Infinitesimal instability}
	\label{sec: infinitesimal instability}

The nonstandard non-Bloch band theory is relevant solely in the presence of symplectic reciprocity in Eq.~(\ref{eq: symplectic symmetry}). Because of this symmetry-protected nature, continuum bands in the symplectic class are fragile against a reciprocity-breaking perturbation, even if it is infinitesimal. As shown in Sec.~\ref{sec: non-Bloch-symplectic; general}, reciprocity forbids the terms proportional to $( \beta_{2M}^{-1} \beta_{2M-1}^{-1} \cdots \beta_{1}^{-1} )^{L}$ for the boundary conditions. However, if we break reciprocity, such terms generally appear and become the leading-order term, and the standard non-Bloch band theory characterizes continuum bands. 

Now, let $\epsilon > 0$ be the degree of the reciprocity-breaking perturbation. Since the leading-order term $\epsilon\,( \beta_{2M}^{-1} \beta_{2M-1}^{-1} \cdots \beta_{1}^{-1} )^{L}$ and the next-to-leading-order term $( \beta_{2M} \beta_{2M-1}^{-1} \cdots \beta_{1}^{-1} )^{L}$ should be comparable to each other for continuum bands, we have
\begin{equation}
\epsilon\,| \beta_{2M}^{-1} \beta_{2M-1}^{-1} \cdots \beta_{1}^{-1} |^{L}
= | \beta_{2M} \beta_{2M-1}^{-1} \cdots \beta_{1}^{-1} |^{L}.
	\label{eq: epsilon - standard}
\end{equation}
Thus, for $L \to \infty$, we indeed have $| \beta_{2M}^{-1} | = | \beta_{2M} |$, i.e., Eq.~(\ref{eq: Yao-Wang-Yokomizo-Murakami-symplectic}). We emphasize that this condition is respected even for infinitesimal but nonzero $\epsilon > 0$. More precisely, the reciprocity-breaking perturbation should at least exceed $\epsilon \sim \mathcal{O}\,( \left| \beta_{2M} \right|^{2L} )$ to satisfy Eq.~(\ref{eq: epsilon - standard}). Notably, the order of the two limits $L \to \infty$ and $\epsilon \to 0$ plays a significant role. If we take $L \to \infty$ first, the standard non-Bloch band theory is relevant even for $\epsilon \to 0$. By contrast, if we take $\epsilon \to 0$ first, the left-hand side of Eq.~(\ref{eq: epsilon - standard}) vanishes and Eq.~(\ref{eq: epsilon - standard}) cannot be satisfied even for $L \to \infty$, resulting in the nonstandard non-Bloch band theory in the symplectic class.

For example, let us add a reciprocity-breaking perturbation to the symplectic Hatano-Nelson model. Physically, such a perturbation can be a magnetic field. In the absence of the perturbation, the characteristic equation is quartic and the four solutions $\beta_{1}, \beta_{2}, \beta_{1}^{-1}, \beta_{2}^{-1}$ come in Kramers pairs, as investigated in Sec.~\ref{sec: symplectic-Hatano-Nelson}. For $\left| g \right| > \left| \Delta \right|$, we have
\begin{equation}
\left| \beta_{1} \right| = \left| \beta_{2} \right|
\neq 1 \neq | \beta_{1}^{-1} | = | \beta_{2}^{-1} |,
	\label{eq: sHN-four}
\end{equation}
which is consistent with the continuum-band condition~(\ref{eq: Yao-Wang-Yokomizo-Murakami-symplectic}) in the symplectic class. In the presence of the perturbation, on the other hand, Eq.~(\ref{eq: Yao-Wang-Yokomizo-Murakami}) of the standard non-Bloch band theory is relevant, as described above. However, Eq.~(\ref{eq: sHN-four}) does not clearly satisfy Eq.~(\ref{eq: Yao-Wang-Yokomizo-Murakami}). As a result, the continuum bands make a dramatic difference. Even the non-Hermitian skin effect can vanish because of such an infinitesimal magnetic field~\cite{OKSS-20}.

\section{Other symmetry classes}
	\label{sec: other symmetry}

\subsection{Symplectic class with additional symmetry}

We have demonstrated that the non-Bloch band theory is altered by symplectic reciprocity in Eq.~(\ref{eq: symplectic symmetry}). In the 38-fold classification of internal-symmetry classes~\cite{KSUS-19}, the simplest symmetry class relevant to this nonstandard non-Bloch band theory is class $\text{AII}^{\dag}$, in which only symplectic reciprocity is present. The nonstandard non-Bloch band theory replaces the standard one also in other symmetry classes, as long as symplectic reciprocity is respected. For example, it is relevant even in the presence of additional sublattice symmetry $\mathcal{S}$ or pseudo-Hermiticity $\eta$. The possible symmetry classes are classified by Ref.~\cite{OKSS-20} on the basis of the relationship between the intrinsic non-Hermitian topology and skin effects. Other than class AII$^{\dag}$, such symmetry classes in one dimension include class DIII$^{\dag}$, class C with $\mathcal{S}_{+}$, class AI with $\eta_{-}$, class CI with $\mathcal{S}_{++}$ or $\eta_{--}$, class BDI with $\mathcal{S}_{+-}$ or $\eta_{-+}$, and class D with $\mathcal{S}_{-}$ (see Tables~S7-S9 in the Supplemental Material of Ref.~\cite{OKSS-20} for details).

It should be noted that certain symmetry leads to real-valued wavenumbers and restores the conventional Bloch band theory. For example, reciprocity without internal degrees of freedom leads to the absence of skin effects and replaces the non-Bloch band theory with the Bloch band theory, as discussed in Sec.~\ref{sec: orthogonal reciprocity}. In a similar manner, certain additional symmetry can replace the nonstandard non-Bloch band theory with the conventional Bloch band theory even in the presence of symplectic reciprocity.

\subsection{Particle-hole symmetry}

Another important internal symmetry is particle-hole symmetry, which is defined by
\begin{equation}
\mathcal{C} H^{T} \mathcal{C}^{-1} = - H,\quad
\mathcal{C}\mathcal{C}^{*} = \pm 1,
\end{equation}
with a unitary matrix $\mathcal{C}$. This symmetry is generally relevant to non-Hermitian superconductors. For $\mathcal{C}\mathcal{C}^{*} = + 1$ ($-1$), non-Hermitian Hamiltonians are defined to belong to class D (C)~\cite{KSUS-19}. In terms of the bulk Hamiltonian $H \left( \beta \right)$, it imposes
\begin{equation}
\mathcal{C} H^{T} \left( \beta \right) \mathcal{C}^{-1} = - H\,( \beta^{-1} ).
\end{equation}
Now, let $E \in \mathbb{C}$ be an eigenenergy of $H \left( \beta \right)$ and $\ket{\phi}$ ($\ket{\chi}$) be the corresponding right (left) eigenstate. Then, we have
\begin{equation}
H\,( \beta^{-1} )\,( \mathcal{C} \ket{\chi}^{*} )
= - E\,( \mathcal{C} \ket{\chi}^{*} ),
\end{equation}
which means that $\mathcal{C} \ket{\chi}^{*}$ is an eigenstate of $H\,( \beta^{-1} )$ with the eigenenergy $-E$. Hence, particle-hole symmetry generally creates a pair of eigenstates with the opposite eigenenergies $\left( E, -E \right)$. This is contrasted with reciprocity, which imposes a constraint on each eigenenergy. 

Still, particle-hole symmetry makes zero energy $E=0$ special and imposes a constraint on zero-energy states. In particular, the zero-energy states do not exhibit skin effects for $\mathcal{C}\mathcal{C}^{*} = + 1$ (class D). To see this, we focus on the characteristic equation $\det \left[ H \left( \beta \right) - E \right] = 0$. Because of particle-hole symmetry, we have
\begin{eqnarray}
\det \left[ H\,( \beta^{-1} ) - E \right] 
&=& \det \left[ -\mathcal{C} H^{T} \left( \beta \right) \mathcal{C}^{-1} - E \right] \nonumber \\
&=& \det \left[ - H \left( \beta \right) - E \right].
\end{eqnarray}
For generic $E \in \mathbb{C}$, this equation does not have direct relationships with the original characteristic equation $\det \left[ H \left( \beta \right) - E \right] = 0$. For $E = 0$, however, we have
\begin{equation}
\det\,[ H\,( \beta^{-1} ) ] 
= \det\,[ H \left( \beta \right) ]
= 0,
\end{equation}
which implies that $\beta^{-1}$ is another solution to the characteristic equation with $E = 0$. Thus, in a similar manner to the orthogonal class discussed in Sec.~\ref{sec: orthogonal reciprocity}, zero-energy states are delocalized and no skin effects occur in the presence of particle-hole symmetry. It should be stressed that this discussion does not necessarily mean delocalization of generic eigenstates with $E \neq 0$ even if the Hamiltonian respects particle-hole symmetry. Furthermore, it is applicable only to zero modes in continuum bands, and Majorana zero modes isolated from continuum bands can be localized.

For $\mathcal{C}\mathcal{C}^{*} = -1$ (class C), on the other hand, $\ket{\phi}$ and $\mathcal{C} \ket{\chi}^{*}$ with $E = 0$ form a Kramers pair, and the above discussions are not applicable in a similar manner to the symplectic class. Consequently, the zero-energy states, if present, should be described by the nonstandard non-Bloch band theory. We note, however, that such zero-energy skin states may be forbidden to appear in continuum bands for a different reason. Actually, in one-dimensional systems in class C, no zero-energy skin state is protected by intrinsic non-Hermitian topology (see Table~S7 in the Supplemental Material of Ref.~\cite{OKSS-20}). The non-Bloch band theory and the skin effects in non-Hermitian superconductors need further study, which we leave for future work.

\subsection{Commutative unitary symmetry and spatial symmetry}

Similar modification of the non-Bloch band theory can arise from symmetry that is not included in the 38-fold internal symmetry. For example, when unitary symmetry that commutes with the Hamiltonian is present, the Hamiltonian is block diagonal in the eigenbasis of the symmetry: $H \left( \beta \right) = \bigoplus_i H_{i} \left( \beta \right)$. The symplectic Hatano-Nelson model discussed in Sec.~\ref{sec: symplectic-Hatano-Nelson} respects such unitary symmetry for $\Delta = 0$. Physically, this means conservation of spin due to the absence of spin-orbit interaction. In this case, $H_{i} \left( \beta \right)$'s do not interact with each other. Consequently, the non-Bloch band theory should be applied not to the original Hamiltonian $H \left( \beta \right)$ but to each subspace $H_{i} \left( \beta \right)$. In a similar manner to the symplectic class discussed in Sec.~\ref{sec: infinitesimal instability}, non-Hermitian systems with commutative unitary symmetry are fragile even against an infinitesimal perturbation, as discussed in Ref.~\cite{Okuma-19}.

Spatial symmetry can also change the non-Bloch band theory. For example, Ref.~\cite{Rui-19} found a reciprocal skin effect in the presence of reflection symmetry. We point out that this reflection-symmetry-protected skin effect should be accompanied by the modification of the standard non-Bloch band theory in a similar manner to the modification due to symplectic reciprocity discussed in this work. Our nonstandard non-Bloch band theory in the symplectic class can further be modified in the presence of such additional symmetry.

\section{Conclusion}
	\label{sec: conclusion}
	
In this work, we have established the non-Bloch band theory of non-Hermitian Hamiltonians in the symplectic class. In contrast to the standard non-Bloch band theory, which describes generic non-Hermitian systems without symmetry, reciprocity in Eq.~(\ref{eq: symplectic symmetry}) leads to Kramers degeneracy and changes the condition for continuum bands. As a consequence of this nonstandard non-Bloch band theory, non-Hermitian skin effects are allowed to occur even in the presence of reciprocity. This contrasts with the orthogonal class, in which reciprocity in Eq.~(\ref{eq: orthogonal symmetry}) forbids the skin effect.

Remarkably, Refs.~\cite{Zhang-19, OKSS-20} identified the origin of the skin effects as non-Hermitian topology that has no counterparts in Hermitian systems. The correspondence between a $\mathbb{Z}$ topological invariant and the non-Bloch band theory was shown. On the basis of this understanding, Ref.~\cite{OKSS-20} further revealed the reciprocity-protected skin effect that is ensured by a $\mathbb{Z}_{2}$ topological invariant. It merits further study to investigate a similar correspondence between this $\mathbb{Z}_{2}$ topological invariant and the nonstandard non-Bloch band theory developed in the present work.

\section*{Acknowledgment}

We thank Ken Shiozaki for helpful discussions. This work was supported by a Grant-in-Aid for Scientific Research on Innovative Areas ``Topological Materials Science" (KAKENHI Grant No.~JP15H05855) from the Japan Society for the Promotion of Science (JSPS), and by JST CREST Grant No.~JPMJCR19T2. K.K. was supported by KAKENHI Grant No.~JP19J21927 from the JSPS. N.O. was supported by KAKENHI Grants No.~JP18J01610 and No.~JP20K14373 from the JSPS. M.S. was supported by KAKENHI Grants No.~JP17H02922 and No.~JP20H00131 from the JSPS.

\medskip
{\it Note added.\,---\,}Recently, we became aware of a related work~\cite{Yi-Yang-20}.

\bibliography{NH-topo}

\end{document}